\documentclass[%
 reprint,
amsmath,amssymb,
pra,
]{revtex4-2}

\usepackage{graphicx}
\usepackage{dcolumn}
\usepackage{bm}
\usepackage{mathcomp}
\usepackage{amsmath}
\usepackage{subfigure}
\usepackage{comment}
\usepackage{hyperref}
\usepackage{color}
\hypersetup{breaklinks=true}

\begin{document}

\preprint{APS/123-QED}

\title{Spin-dependent metastable He ($2^3S$) atom scattering from ferromagnetic surfaces: Potential application to polarized-gas production}\

\author{Haruka Maruyama$^1$}
\author{Mitsunori Kurahashi$^2$}
\email{KURAHASHI.Mitsunori@nims.go.jp}
\author{Kanta Asakawa$^1$}
\author{Atsushi Hatakeyama$^1$}
\email{hatakeya@cc.tuat.ac.jp}
\affiliation{$^1$Department of Applied Physics, Tokyo University of Agriculture and Technology, Koganei, Tokyo 184-8588, Japan}
\affiliation{$^2$National Institute for Materials Science, 1-2-1 Sengen, Tsukuba, Ibaraki 305-0047, Japan}

\date{\today}

\begin{abstract}
A spin-polarized triplet metastable helium (He*) beam has been used as a probe for surface magnetism, but changes in the spin state during scattering from a surface remain unclear. In the present study, we explored this issue by constructing an apparatus that allows us to direct a spin-polarized He* beam to a surface and measure the spin polarization of He* scattered from the surface. Magnetic hexapoles were used for both the beam polarization and the spin analysis. The results of the spin-dependent He* scattering experiments on clean Fe$_3$O$_4$(100), H-terminated Fe$_3$O$_4$(100), benzene-adsorbed Fe$_3$O$_4$(100), and non-magnetic Cu(100) surfaces indicated that although the spin direction of He* was mostly preserved during scattering from these surfaces, spin-flop scattering of surviving He* occurred with a probability up to approximately 0.1. Our results showed that the survival probability was higher when the spins of He* and the Fe$_3$O$_4$(100) film were parallel, which can be understood based on the lower He* resonance ionization rate for this spin orientation. Based on our findings, we estimate that a non-polarized He* gas becomes 10\% spin-polarized after a single collision with a clean Fe$_3$O$_4$(100) surface.
\end{abstract}

\maketitle

\section{\label{sec:introduction}introduction}
Spin-polarized gaseous atoms have many research applications, including those involving atomic magnetometers~\cite{doi:10.1063/1.3491215}, neutron spin filters~\cite{OKUDAIRA2020164301}, and magnetic resonance imaging of the lung~\cite{https://doi.org/10.1002/jmri.20154}. Spin-polarized gases are often produced and used in containers, but spin depolarization due to collisions with container walls is detrimental to these applications. The effect of surface magnetism on depolarization, considered a key factor, has not yet been clarified due to a lack of experimental evidence. 

In the field of condensed matter physics, the collision of spin-polarized particle beams with magnetic surfaces has been utilized to study surface magnetism. A spin polarized triplet metastable helium [He*($2^3S$)] beam, for which the magnetic sublevel is specified as $M_S=+1$ or $-1$, has been used to investigate the spin polarization of the topmost surface of a solid~\cite{PhysRevLett.52.380}. Most He* atoms de-excite on the vacuum side of the surfaces and induce the emission of surface electrons, for which the kinetic energy distribution is measured in spin-polarized metastable deexcitation spectroscopy (SPMDS)~\cite{PhysRevLett.52.380,PhysRevB.45.3674,GETZLAFF19951404,PhysRevB.54.14758,PhysRevB.81.193402}. However, a fraction of impinging He* survive the collision and are scattered from the surface. Previous studies have indicated that the survival probability ranges from $10^{-3}$ to $10^{-6}$, depending on the surface electronic states~\cite{CONRAD1982281,CONRAD198298,FOUQUET1998140} and the spin orientation of the incident He* with respect to the surface spin~\cite{PhysRevLett.91.267203}. However, it has remained unclear whether the spin state of He* changes during scattering. Experimental verification is required, given the efficiency of spin depolarization in alkali atom scattering~\cite{PhysRevA.98.042709}, as well as phenomena that include the spin-flip scattering of polarized neutrons~\cite{PhysRev.181.920} and nuclear spin conversion of ortho-H$_2$ on surfaces~\cite{FUKUTANI2013279}. Such information can also provide insight into the possibility of producing a spin-polarized gas via scattering from a ferromagnetic target.

In the present study, to determine how the spin state of He* changes during scattering from a surface, we constructed an apparatus that allows us to direct a spin polarized triplet He* beam to a surface and measure the spin polarization of scattered He*. To reveal the effects of surface magnetism and surface termination, we conducted spin-dependent He* scattering experiments with clean Fe$_3$O$_4$(100), H-terminated Fe$_3$O$_4$(100), benzene-adsorbed Fe$_3$O$_4$(100), and non-magnetic Cu(100) surfaces. The results indicated that although the He* spin direction was mostly preserved during scattering, spin-flop scattering of survived He* occurred with a probability up to approximately 0.1. We also found that the He* survival probability was higher when spins of He* and the Fe$_3$O$_4$(100) film were parallel, which can be understood based on the lower He* resonance ionization (RI) rate for this spin orientation. From these findings, we estimate that a non-polarized He* gas becomes 10\% spin-polarized after a single collision with a clean Fe$_3$O$_4$(100) surface.

This paper is organized as follows. In Sec. \ref{sec:apparatus}, we describe the experimental apparatus. The measurement of the spin polarization of survived He*, together with the measurements of the survived He* intensity and SPMDS spectra conducted for comparison, are presented in Sec. \ref{sec:experiments}. We discuss the spin flop of scattered He* and possible polarization achieved by scattering of a non-polarized He* beam on an Fe$_3$O$_4$(100) surface in Sec. \ref{sec:discussion}. The paper is concluded in Sec. \ref{sec:conclusions}.

\section{\label{sec:apparatus}Experimental Apparatus}
\subsection{\label{subsec:beam}Spin polarized He* beam}
Figure \ref{fig:souti} shows a schematic diagram of the experimental apparatus. All experiments were conducted in an ultrahigh vacuum chamber. The main part of the apparatus was the same as that used in our previous study~\cite{Kurahashi:2021vj}. A hexapole-type spin polarization analyzer for scattered He*, which will be described in Sec.~\ref{subsec:analysis}, was added for the present study. He* atoms were generated by cold DC discharge and skimmed to produce the He* beam. He*(2$^3S$) atoms with a magnetic quantum number of $M_S=+1$ were focused with a hexapole magnet (the polarizer hexapole) ~\cite{Baum:1988va} to pass through an aperture located downstream. He*(2$^1S$) and photons produced by discharge were removed by the center stop. The photons could not be completely blocked by the center stop and resulted in the background of the scattered He* signal. The background caused by scattering of the residual photons was not negligible, given that the He* survival probability was low. A leak valve attached to the hexapole chamber allowed us to introduce He gas to quench the He* atoms in the beam without changing the photon intensity. This enabled us to derive the photon background. $M_S$ was defined with respect to the local magnetic field; its polarity could be flipped with a spin flipper that consisted of two longitudinal coils and a transverse coil. The beam was operated in two different modes of the spin flipper, the non-flip mode and the flip mode. Stern-Gerlach analysis of the He* beam generated with a similar polarizer hexapole and a flipper~\cite{doi:10.1063/1.2949385} indicated that the $M_S = +1$ concentration in the nonflip mode was nearly 100\%, whereas 90\% of He* atoms were in the $M_S = -1$ state and 10\%($=\beta$) were in the $M_S = 0$ state in the flip mode. After the spin flipper, the He* spin state was adiabatically transported to the sample position, which was checked by analyzing the beam spin polarization using the method described in our previous paper \cite{doi:10.1063/5.0031903}. The He* spin direction at the sample was controlled by applying a defining magnetic field ($H$) of 0.15 G using three-axis coils located behind the sample. 

\subsection{Spin polarization analysis of scattered He*}
\label{subsec:analysis}
\begin{figure}
\centering
\includegraphics[width=.45\textwidth]{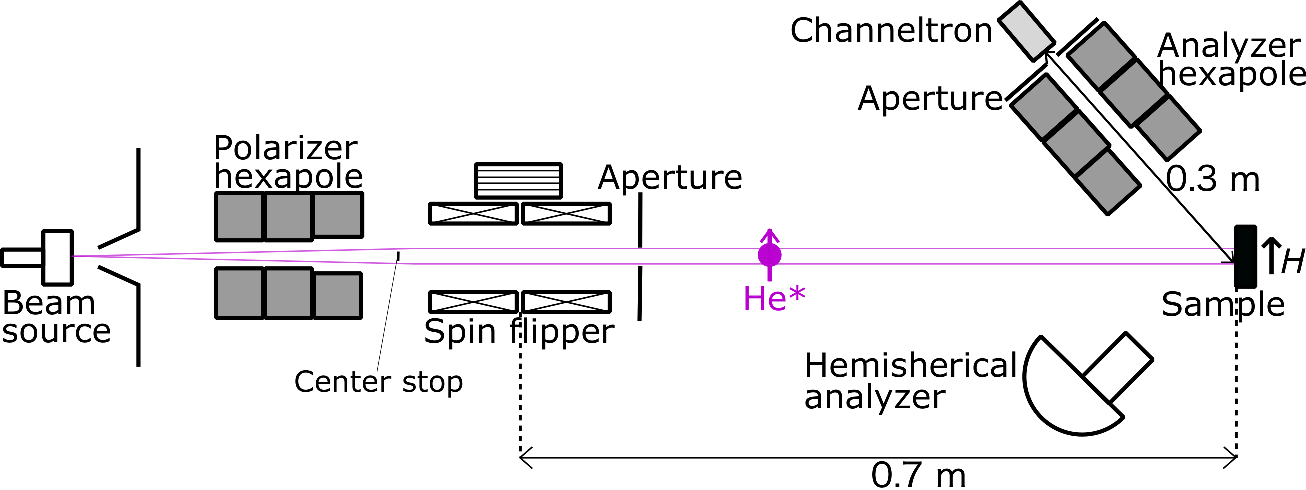}
\caption{\label{fig:souti} Schematic diagram of the experimental apparatus. The distances from the center of the spin flipper to the sample, and from the sample to the entrance of the channeltron, were 0.7 and 0.3~m, respectively. The diameter of the apertures was 2~mm.}
\end{figure}

The spin polarization of the scattered He* atoms was analyzed with a combination of a hexapole, an aperture, and a channeltron, as illustrated in Fig.~\ref{fig:souti}. The structure of this spin analyzer is the same as that used for the He* beam polarization analysis reported previously~\cite{doi:10.1063/5.0031903}. A hexapole with a structure identical to the polarizer hexapole was employed. This assembly was mounted on a tilt stage that allowed us to optimize the detection geometry. Because of the focusing (defocusing) by the hexapole, the steric angular range of scattered He* atoms that were collected, magnetically deflected and finally passed through the aperture before the channeltron was larger (smaller) for $M_S=+1$ $(-1)$. The efficiency for the analyzer for collecting He* ($M_S=+1$) was therefore much higher than that for $M_S=-1$. The collection efficiency for He* ($M_S = 0$), $\gamma$, was estimated to be 0.12 using the method described in \footnote{We placed the same spin analyzer in the beamline and measured the change in the detector signal when changing the spin flipper mode. The signal intensity for the flip mode was 0.012 of that for the non-flip mode. Here, 10\% of He* in the beam for the flip mode is in the $M_S=0$ state and 90\% is in the $M_S=-1$ state while the detection efficiency for $M_S=-1$ is negligibly low. Therefore, we can estimate the detection efficiency for $M_S=0$ to be 0.12 of that for $M_S=+1$.}. In the present study, the He* beam impinged along the surface normal, and He* atoms scattered to 45$^{\circ}$ relative to the normal were detected.

The SPMDS measurement was performed by measuring the energy distribution of electrons ejected by He* using a hemispherical analyzer attached at 45$^{\circ}$ to the beam axis. In the present experiment, the intensities of electrons ejected and He* scattered from the same beam spot on the sample were measured simultaneously. Here, the beam size on the sample was 2-3 mm in diameter.

\subsection{Spin configurations in the scattering}
Four nonequivalent spin configurations exist, as shown in Fig.~\ref{fig:scat}; these were realized by controlling the defining field direction at the sample and the spin flipper mode. For example, the nonflip (N) and parallel (P) mode [Fig.~\ref{fig:scat}(a)] corresponds to the collision in which the He* spin direction remains unchanged when He* impinges on the surface with its spin parallel to the majority spin of the sample ($S_{M}$). This case is realized by setting the flipper to the N mode and directing $H$ parallel to $S_{M}$. In the flip (F) and antiparallel (A) mode [Fig.~\ref{fig:scat}(d)], the spin flipper is in the F mode and $H$ is parallel to $S_{M}$.

\subsection{Time-of-flight analysis}
\begin{figure}
\centering
\includegraphics{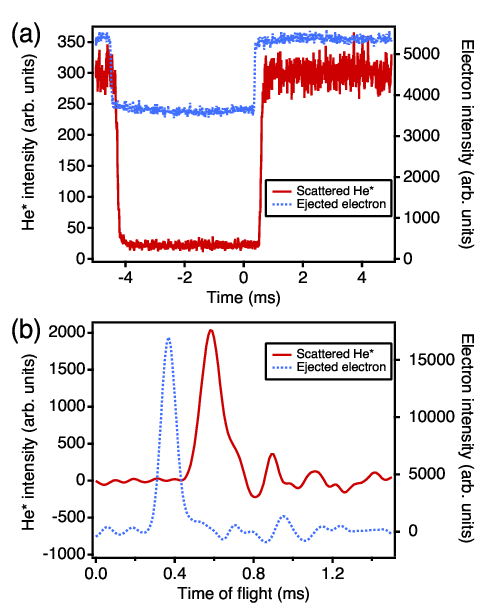}
\caption{\label{fig:tof} (a) Time-resolved intensities of He* scattered from a clean Fe$_3$O$_4$(100) surface and accompanying ejected electrons. The spin configuration was changed at $-5$ and 0~ms by switching the mode of the spin flipper as described in the text. (b) Time-of-flight spectra derived by differentiating the profiles of (a).}
\end{figure}

Measurement of the time-resolved intensities of scattered He* and ejected electrons while switching the flipper mode allowed us to conduct a time-of-flight (TOF) analysis of incident and scattered He*. Figure~\ref{fig:tof}(a) shows the time-resolved intensities of He* scattered from a clean Fe$_3$O$_4$(100) surface and accompanying ejected electrons. The spin configuration was changed from the N and P mode to the F and A mode at $-5$~ms, and then back to the N and P mode at 0~ms by switching the mode of the spin flipper. Here, the change in the spin flipper mode causes the change in the He* and ejected electron intensities because these intensities depend on the spin configurations. The profile of the ejected electron (broken line) and scattered He* (solid line) [Fig.~\ref{fig:tof}(a)] exhibits time delay from the switching signal. Here, we define $t_1$ as the TOF of the incident He* from the center of the spin flipper to the sample, and $t_2$ as the TOF of the scattered He* from the sample to the detector of the spin analyzer. The time delay of the ejected electron signal is nearly equal to $t_1$ because the electron's TOF is negligibly short. The time delay of scattered He* is equal to $t_1+t_2$. Figure~\ref{fig:tof}(b) shows the numerical derivatives of the time profiles in Fig.~\ref{fig:tof}(a). The position of the peaks in Fig.~\ref{fig:tof}(b) allows us to evaluate the kinetic energies of the incident and scattered He* atoms, which were 76 and 47~meV, respectively. The following two factors might cause the observed energy difference between the incident and scattered He*. The first is the energy loss during the collision. Previous He* scattering study on Cu and NiO surfaces~\cite{FOUQUET1998140} showed energy losses of 20-30~meV, which are similar to the energy difference observed here. Considering that masses of the target atoms on Fe$_3$O$_4$ are similar to those for NiO, the observed similar energy loss would be reasonable. However, the origin of the energy loss is not clear from the present results. The second is the difference in the He* pass energy in the polarizer and analyzer hexapoles. Although hexapoles with an identical structure were used for the polarizer and analyzer, the kinetic energy of He* which was focused by the magnetic lens optics of the analyzer would be lower due to its shorter focal distance. 

\subsection{Sample preparation}
MgO(100) substrates were cleaned by degassing at 873~K under ultrahigh vacuum conditions and then annealing at 873~K for 15 min in an O$_2$ atmosphere. This annealing process completely removed carbon contamination, as verified by x-ray photoelectron spectroscopy. A 22-nm-thick Fe$_3$O$_4$ film was grown on the cleaned MgO(100) substrate at 503~K by evaporating iron in an O$_2$ atmosphere with a deposition rate of 0.02 \AA/s for 180 min. Low-energy electron diffraction (LEED) clearly indicated $(\sqrt{2}\times\sqrt{2})R45^\circ$ reconstruction of the Fe$_3$O$_4$ surface, which agreed well with a previous report~\cite{PhysRevB.72.104436}. The Fe$_3$O$_4$ film was magnetized along the in-plane easy axis by a current pulse through a coil.
To prepare H-terminated Fe$_3$O$_4$(100) surfaces, atomic H obtained by dissociating H$_2$ molecules on a heated tungsten filament was adsorbed on clean Fe$_3$O$_4$(100) films. The H-termination was confirmed by observing the LEED pattern reflecting the $(1 \times 1)$ structure~\cite{PhysRevB.81.193402, Pratt_2011}
In the experiment on benzene-adsorbed surfaces, the H-terminated Fe$_3$O$_4$ substrate was cooled to 117~K before exposure to purified benzene vapor.
A clean Cu(100) surface was prepared by repeating 1~keV Ar$^+$ sputtering and annealing at 863~K. 
\begin{figure}
\centering
\includegraphics{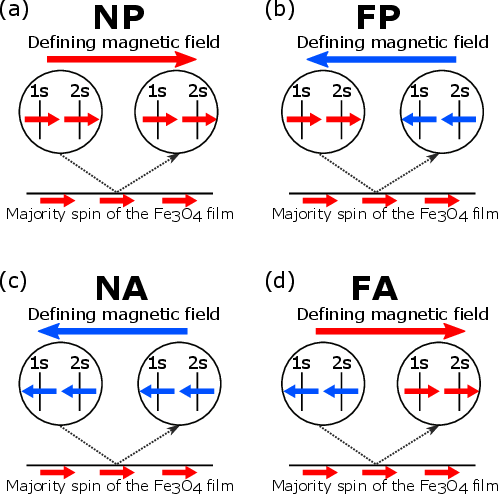}
\caption{\label{fig:scat} Schematic illustration of the incident and measured spin states in the four spin modes. (a) Non-flip and parallel mode. (b) Flip and parallel mode. (c) Non-flip and antiparallel mode. (d) Flip and antiparallel mode.}
\end{figure}

\section{\label{sec:experiments}Results}
\begin{figure}
\centering
\includegraphics{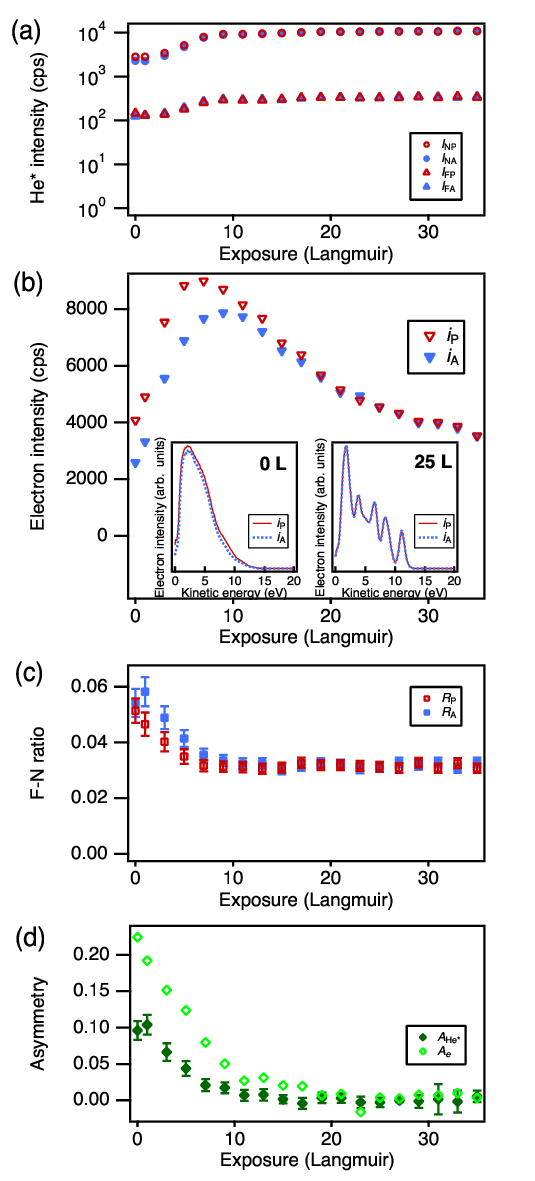}
\caption{(a) Benzene exposure dependence of scattered He* intensities for the H-terminated Fe$_3$O$_4$(100) surface, (b) the intensities of ejected electrons at 10~eV, (c) F-N ratios of scattered He* intensities, and (d) spin asymmetries of scattered He* and ejected electrons. Statistical errors are shown in (c) and (d). Insets in (b): SPMDS spectra}
\label{fig:exposure}
\end{figure}
Figure~\ref{fig:exposure}(a) shows the benzene exposure dependence of the intensity of He* scattered from the H-terminated Fe$_3$O$_4$(100) measured at the four spin configurations illustrated in Fig.~\ref{fig:scat}. $I\rm{_{NP}}$, $I\rm{_{NA}}$, $I\rm{_{FP}}$, and $I\rm{_{FA}}$ represent the intensities of scattered He* in N and P mode, N and A mode, F and P mode, and F and A mode, respectively. The background due to the scattered photons described in Sec.~\ref{subsec:beam} was subtracted from the He* intensity. The results show that intensity increases largely with benzene exposure at 0-10 L (1~L= 1.3$\times$10$^{-4}$~Pa$\cdot$s.). The difference in intensity between the P and A modes is also found in this exposure range. Figure~\ref{fig:exposure}(b) shows the ejected electron intensities in the P and A modes, which are represented as $i_{\mathrm{P}}$ and $i_{\mathrm{A}}$, respectively, monitored at an energy of 10 eV. The ejected electron intensities show clear dependencies on the benzene exposure and spin orientation at 0-10 L.

These behaviors are associa ted with the variation in benzene coverage and the He* deexcitation mechanism on the surface. On the unexposed H-terminated Fe$_3$O$_4$(100) surface, He* decays via RI, followed by Auger neutralization (AN). The broad feature of the SPMDS spectra at 0~L shown in the inset of Fig.~\ref{fig:exposure}(b) confirms this. Benzene adsorbs molecularly on the surface. Its monolayer forms at approximately 5~L, a multilayer island begins to form at 5~L, and the whole surface is covered with a multilayer at approximately 10~L ~\cite{Pratt_2011, KURAHASHI200521}. The SPMDS spectra at 25~L [Fig.~\ref{fig:exposure}(b) inset] showing photoemission-like features reflect the multilayer formation, as well as He* deexcitation via the Auger deexcitation (AD) process. Here, because the RI is more efficient than AD ~\cite{SESSELMANN198417}, the intensity of surviving He* is higher at larger exposures [Fig.~\ref{fig:exposure}(a)]. The clean and submonolayer covered surfaces are spin polarized, causing the spin dependence in surviving He* and the ejected electron intensities at 0-10~L. The situation that both the ejected electron intensity and the survival probability of He* are higher for the spin parallel configuration can be understood as follows. The ejected electron intensity is proportional to the product of the He* deexcitation probability and the electron yield per He*. Since the deexcitation probability is equal to one minus  survival probability, it is almost unity because the He* survival probability would be very low ($< 10^{-3}$). Therefore, although the survival probability was higher for the spin parallel configuration, its effect on the deexcitation probability is negligibly small.  Therefore, the higher survival probability does not conflict with the higher ejected electron intensity.

\begin{figure}
\centering
\includegraphics[width=.45\textwidth]{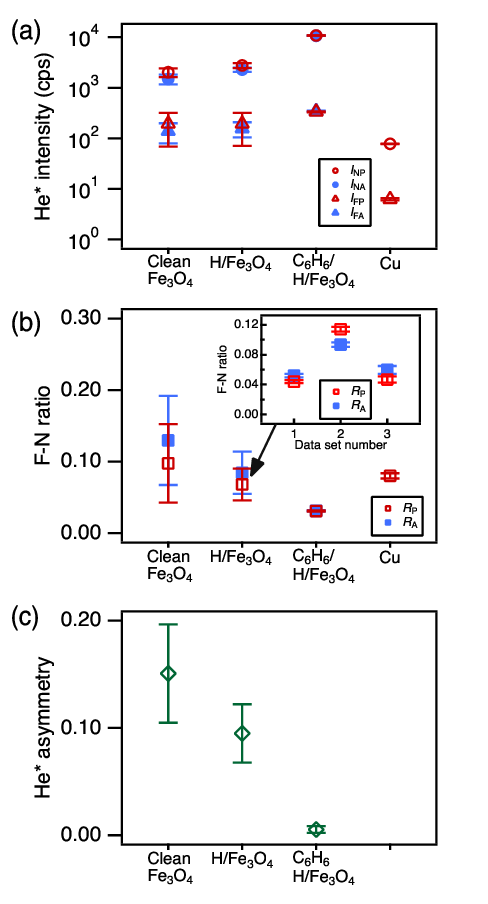}
\caption{(a) The intensities of He* scattered from clean Fe$_3$O$_4$(100), H-terminated Fe$_3$O$_4$(100), multilayered benzene on H-terminated Fe$_3$O$_4$(100), and Cu(100). (b) F-N ratios of intensities of scattered He*. (c) Spin asymmetry of scattered He*. For clean Fe$_3$O$_4$(100) and H-terminated Fe$_3$O$_4$(100), the means of multiple measurements are shown with standard deviations as error bars. Statistical errors are shown for the results of benzene and Cu(100). Inset in (b): the F-N ratio of each measurement set for the H-terminated Fe$_3$O$_4$ surface.}
\label{fig:IRA}
\end{figure}

Figure~\ref{fig:exposure}(c) shows the intensity ratio of the scattered He* between spin configurations (F-N ratio). The F-N ratio roughly represents the spin-flopping probability, which will be discussed in more detail in Sec.~\ref{subsec:spinflop}. $R\mathrm{_P}$ ($R\mathrm{_A}$) corresponds to the spin-flop scattering where the incident He* spin is parallel (antiparallel) to $S_M$.  
\begin{equation}
R\mathrm{_P}=\frac{I\mathrm{_{FP}}}{I\mathrm{_{NP}}},
\label{eq:two}
\end{equation}
\begin{equation}
R\mathrm{_{A}}=\frac{I\mathrm{_{FA}}}{I\mathrm{_{NA}}}.
\end{equation}
$R\mathrm{_P}$ ($R\mathrm{_A}$) corresponds to the spin-flop scattering where the incident He* spin is parallel (antiparallel) to $S_M$. The very small values of these ratios indicate that the He* spin direction is mostly preserved during scattering. However, these ratios also show clear exposure dependence [Fig.~\ref{fig:exposure}(c)], indicating that a transition between different magnetic sublevels of He* happens during scattering from the surface. Small differences are discernible between $R\mathrm{_P}$ and $R\mathrm{_A}$ at low exposures; however, these differences have some uncertainty, based on the results for different runs that will be shown later [Fig.~\ref{fig:IRA}(b) inset]. 

As shown in Fig.~\ref{fig:exposure}(a), the intensity of scattered He* exhibits a clear dependence on the spin orientation between incident He* and $S_{M}$. To see these differences quantitatively, the spin asymmetries of scattered He* intensities are defined as
\begin{equation}
A\mathrm{_{He*}}=\frac{I\mathrm{_{NP}}-I\mathrm{_{NA}}}{I\mathrm{_{NP}}+ I\mathrm{_{NA}}}.
\label{eq:three}
\end{equation}
The spin asymmetry roughly represents the difference in spin-preserving survival probability between the P and A modes. They will be discussed in more detail in Sec.~\ref{sec:discussion}. Figure~\ref{fig:exposure}(d) shows the spin asymmetries in intensities of He* and the electrons.
The asymmetry in intensity of the ejected electrons at 10~eV $A\rm{_e}$ is defined as follows:
\begin{equation}
A\mathrm{_e}=\frac{ i\mathrm{_P}- i\mathrm{_{A}}}{ i\mathrm{_P}+i\mathrm{_{A}}}.
\label{eq:one}
\end{equation}
As the benzene exposure increased, $A\rm{_{He*}}$ and $A\rm{_e}$ decreased and became zero at approximately 20~L. This exposure dependence is reasonable, given that the topmost surface of the multilayer benzene on H-terminated Fe$_3$O$_4$ is not spin polarized. 

Figure~\ref{fig:IRA}(a) shows a summary of the scattered He* intensities in the four spin modes for clean Fe$_3$O$_4$(100), H-terminated Fe$_3$O$_4$(100), multilayered benzene on H-terminated Fe$_3$O$_4$(100), and Cu(100). Comparing intensities in the same spin modes, we found that those for clean and H-terminated Fe$_3$O$_4$(100) were lower than those for benzene by factors of 4 and 3, respectively, whereas the intensity for Cu was lower by two orders of magnitude. These differences can be explained by the fact that He* decays via the RI, followed by AN on clean and H-terminated Fe$_3$O$_4$(100) and Cu(100), whereas it decays via the AD process on the benzene multilayer~\cite{PhysRevB.81.193402, CONRAD198298} . Moreover, considering the results of previous He* scattering studies~\cite{CONRAD1982281,CONRAD198298,FOUQUET1998140}, the survival probability is expected to be higher for oxidized metal surfaces than clean metallic surfaces. This is because the density of states (DOS) near the Fermi level ($E\rm{_F}$), which determines the He* RI rate, on oxidized metallic surfaces would be lower than on clean metallic surfaces.

Figure ~\ref{fig:IRA}(b) summarizes the F-N ratios ($R\rm{_P}$ and $R\rm{_{A}}$), which reflect the efficiency of the spin-flop scattering (as determined for the four surfaces). The values of F-N ratio derived from the data taken at different experimental runs are plotted in the inset with statistical error bars. The averaged values of these data points are plotted for clean and H-terminated Fe$_3$O$_4$ surfaces in the main part of Fig.~\ref{fig:IRA}(b). The error bars shown include the distribution of these data points. The large distribution of the F-N ratio for different runs might originate from the fact that the He* scattering seems to be sensitive to surface defects and/or the adsorption of minute impurity species contained in the He* beam. Better control of such parameters would allow us to discuss more details of the spin-dependent He* scattering in the future. The results show that (i) the F-N ratio for clean Fe$_3$O$_4$(100) is higher than that for  H-terminated Fe$_3$O$_4$(100), although spin polarization at $E\rm{_F}$ is higher on the H-terminated surface~\cite{PhysRevB.81.193402}. (ii) A difference between $R\rm{_P}$ and $R\rm{_{A}}$ is discernible for the Fe$_3$O$_4$ surfaces, but is less clear when considering the results of multiple runs. (iii) The F-N ratio determined for a multilayer benzene-covered surface is clearly lower than that for other surfaces, and (iv) the F-N ratios for the magnetic Fe$_3$O$_4$ and nonmagnetic Cu(100) surfaces do not differ significantly.

Figure~\ref{fig:IRA}(c) summarizes the $A\rm{_{He*}}$ values, which reflect the spin orientation dependence of the He* survival probability.  The results indicate that $A\rm{_{He*}}$ for the clean Fe$_3$O$_4$ surface is higher than that for the H-terminated surface, although the spin polarization at $E\rm{_F}$ is lower on the clean Fe$_3$O$_4$(100) surface. The much lower $A\rm{_{He*}}$ on multilayer benzene reflects the  fact that the topmost layer of multilayer benzene is nearly unpolarized.
\begin{figure}
\centering
\includegraphics{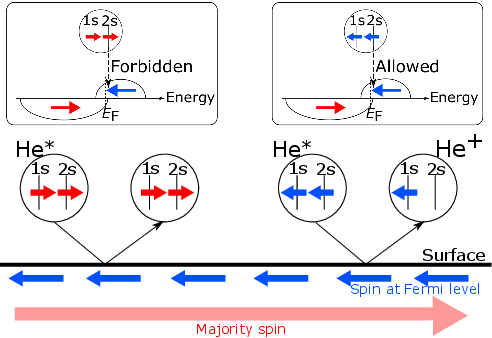}
\caption{\label{fig:pontie} Schematic illustration of the processes of He* survival and deexcitation on the surface of Fe$_3$O$_4$(100).}
\end{figure}

\section{\label{sec:discussion}Discussion}
\subsection{\label{subsec:spinflop}Spin-flop scattering of He*}

First, we discuss the spin-flop scattering of He*. To characterize the dependence of the scattering processes on the spin directions, we consider the transition probability matrix $(a_{ij})$ and survival probability $b_j$. Incident He* atoms with their spins parallel, antiparallel, and perpendicular to $S_{M}$, survive at probabilities of $b_{\mathrm{P}}, b_{\mathrm{A}}$, and $ b_{\mathrm{0}}$, respectively. For surviving He* atoms, the matrix element $a_{ij}$ represents the transition probability from spin directions $j$ to $i$, where $j=$P, A, and 0 represent the spin direction of the He* atom parallel, antiparallel, and perpendicular to $S_{M}$ before scattering, respectively, and $i=$P, A, and $0$ represent the spin direction after scattering. If the scattering completely randomizes the spin states, all elements of the transition probability matrix are equal to 1/3. If the scattering does not change the spin states at all, the diagonal elements of the transition probability matrix are equal to 1, whereas the others are 0. We assumed that our investigated scattering processes should be somewhere between these two extreme cases. 

Using these transition probabilities, together with experimental parameters describing the efficiencies of the spin flipper and detector, $\beta= 0.10$ and $\gamma= 0.12$, respectively, $I_{\rm{NP}}$, $I_{\rm{NA}}$, $I_{\rm{FP}}$, and $I_{\rm{FA}}$ can be expressed as follows:
\begin{equation}
I_{\mathrm{NP}}=(a_{\mathrm{PP}} +\gamma a_{\mathrm{0P}})b_{\mathrm{P}}I,
\label{eq:dis1}
\end{equation}
\begin{equation}
I_{\mathrm{NA}}=(a_{\mathrm{AA}} +\gamma a_{\mathrm{0A}})b_{\mathrm{A}}I,
\label{eq:dis2}
\end{equation}
\begin{equation}
I_{\mathrm{FP}}= \\
\{(1-\beta)a_{\mathrm{AP}} b_{\mathrm{P}}+\beta a_{\mathrm{A0}} b_{\mathrm{0}}+\gamma((1-\beta) a_{\mathrm{0P}} b_{\mathrm{P}}+\beta a_{\mathrm{00}} b_{\mathrm{0}}]\}I\\,
\label{eq:dis3}
\end{equation}
\begin{equation}
I_{\mathrm{FA}}= \\
\{(1-\beta)a_{\mathrm{PA}} b_{\mathrm{A}}+\beta a_{\mathrm{P0}} b_{\mathrm{0}}+\gamma((1-\beta) a_{\mathrm{0A}} b_{\mathrm{A}}+\beta a_{\mathrm{00}} b_{\mathrm{0}}]\}I\\.
\label{eq:dis4}
\end{equation} 
Here, $I$ is the incident He* intensity. The meaning of the above expressions, for example, that of $I_{\mathrm{FP}}$, is as follows. The He* beam in the F mode contained $1-\beta=90$\% He* atoms in the $M_S=-1$ state and $\beta=10$\% in the $M_S=0$ state. The impinging He* atoms in the $M_S=-1$ and $M_S=0$ states survived at probabilities of $b_{\mathrm{P}}$ and $b_{\mathrm{0}}$, respectively. We detected scattered He* atoms that changed their magnetic substates from $M_S=-1$ to $M_S=+1$ at a probability of $a_{\mathrm{AP}}$, from $M_S=0$ to $M_S=+1$ at a probability of $a_{\mathrm{A0}}$, and from $M_S=-1$ to $M_S=0$ at a probability of $a_{\mathrm{0P}}$; they retained their magnetic substate in the $M_S=0$ state at a probability of $a_{\mathrm{00}}$. The detection efficiency of scattered He* atoms in the $M_S=0$ state was $\gamma=0.12$, compared with the efficiency in the $M_S=+1$ state. 

Using Eqs. (\ref{eq:dis1}), (\ref{eq:dis2}), (\ref{eq:dis3}), and (\ref{eq:dis4}), the F-N ratios are expressed as
\begin{equation}
\begin{split}
R\mathrm{_P}&=\frac{I\mathrm{_{FP}}}{I\mathrm{_{NP}}} \\
&=\frac{a_{\mathrm{AP}}+\beta(a_{\mathrm{A0}}\frac{b_{\mathrm{0}}}{ b_{\mathrm{P}}}-a_{\mathrm{AP}}) +\gamma[a_{\mathrm{0P}}+\beta(a_{\mathrm{00}} \frac{b_{\mathrm{0}}}{ b_{\mathrm{P}}}-a_{\mathrm{0P}})]}{a_{\mathrm{PP}} +\gamma a_{\mathrm{0P}}}\\
\label{eq:nine}
\end{split}
\end{equation}
\begin{equation}
\begin{split}
R\mathrm{_{A}}&=\frac{I\mathrm{_{FA}}}{I\mathrm{_{NA}}} \\
&=\frac{a_{\mathrm{PA}}+\beta(a_{\mathrm{P0}} \frac{b_{\mathrm{0}}}{ b_{\mathrm{A}}}-a_{\mathrm{PA}}) +\gamma[a_{\mathrm{0A}}+\beta(a_{\mathrm{00}} \frac{b_{\mathrm{0}}}{ b_{\mathrm{A}}}-a_{\mathrm{0A}})]}{a_{\mathrm{AA}} +\gamma a_{\mathrm{0A}}} \\
\label{eq:ten}
\end{split}
\end{equation}

For the benzene and Cu surfaces, which are not spin-polarized, the survival probabilities should be spin-independent. Thus, in the case of perfect spin-randomizing scattering, 
\begin{gather}
R\mathrm{_P}=R\mathrm{_{A}}=1,
\end{gather}
and in the case of perfect spin-preserving scattering, 
\begin{gather}
R\mathrm{_P}= R\mathrm{_{A}}=\gamma\beta = 0.012.
\end{gather}
The experimental results showed that the F-N ratios were 0.03 and 0.08 for benzene and Cu, respectively, which are larger than 0.012. This implies that spin flopping occurred (i.e., the off-diagonal elements of the transition matrix were nonzero). The F-N ratios were even larger for clean Fe$_3$O$_4$ and H-terminated Fe$_3$O$_4$. If we assume perfect spin-preserving (no spin-flopping) scattering with spin-dependent survival probabilities, 
$R\mathrm{_P}=\gamma\beta b_{\mathrm{0}}/ b_{\mathrm{P}}$ and $R\mathrm{_{A}}=\gamma\beta b_{\mathrm{0}}/ b_{\mathrm{A}}$.
However, it is unrealistic for $b_{\mathrm{0}}/ b_{\mathrm{P}}$ and $b_{\mathrm{0}}/ b_{\mathrm{A}}$ to be large, approximately 10, to explain the observed F-N ratios of approximately 0.1, because it has been shown that $b_{\mathrm{A}}/ b_{\mathrm{P}}$ is close to 1 (e.g., 0.85 for Fe surfaces)~\cite{PhysRevLett.91.267203}. Thus, the assumption of no spin-flopping scattering was incorrect; the spin-flopping scattering occurred for Fe$_3$O$_4$ and H-terminated Fe$_3$O$_4$ surfaces, as well as for benzene and Cu surfaces, and the larger F-N ratios can be attributed to the larger spin flopping rates.

The spin configuration dependence of the F-N ratio may be difficult to discuss based on the present results, as no clear difference was found in the F-N ratio between the P and A modes [Fig.~\ref{fig:IRA} (b)].

\subsection{\label{subsec:asymmetry}Spin asymmetry of scattered He*}
Figure~\ref{fig:IRA}(c) shows that the intensity of He* scattered from clean and H-terminated Fe$_3$O$_4$(100) depends on the spin configuration. The spin asymmetry, which is defined as Eq.~(\ref{eq:three}), can be approximated using Eqs. (\ref{eq:dis1}) and (\ref{eq:dis2}), as follows:
\begin{equation}
A_{\mathrm{He*}}\simeq \frac{a_{\mathrm{PP}}b_{\mathrm{P}}- a_{\mathrm{AA}} b_{\mathrm{A}}}{ a_{\mathrm{PP}} b_{\mathrm{P}}+ a_{\mathrm{AA}} b_{\mathrm{A}}}, 
\end{equation}
where we neglect the matrix elements multiplied by the small factor $\gamma$.
Here, given that $a_{\mathrm{PP}}\sim a_{\mathrm{AA}}$, $A_{\mathrm{He*}}$ is expressed as follows:
\begin{equation}
A_{\mathrm{He*}}\simeq \frac{b_{\mathrm{P}}- b_{\mathrm{A}}}{b_{\mathrm{P}}+ b_{\mathrm{A}}}.
\end{equation}
$A_{\mathrm{He*}}>0$ for clean and H-terminated Fe$_3$O$_4$(100) implies that the survival probability is higher when the He* spin is parallel to $S_{M}$. Similarly to the case of He* scattering from Fe~\cite{PhysRevLett.91.267203}, this finding can be understood based on the spin dependence in the RI rate. Figure \ref{fig:pontie} illustrates the DOS of the Fe$_3$O$_4$(100) surface~\cite{PhysRevB.81.193402}. When He* approaches the surface, its 2s state overlaps with a surface unoccupied state with the same spin, causing the RI. Because the spin-down DOS is higher than the spin-up DOS near $E\rm{_F}$ on Fe$_3$O$_4$(100), the RI rate is higher and the survival probability is lower when the He* spin is antiparallel to $S_{M}$. This explains the positive $A_{\mathrm{He*}}$. 

Figure~\ref{fig:IRA}(c) shows a tendency that $A_{\mathrm{He*}}$ is higher for clean Fe$_3$O$_4$(100) than for its H-terminated surface. The difference in the He* RI rate on Fe and O atoms on the two surfaces needs to be considered to discuss this result. First, the RI rates on Fe atoms on clean and H-terminated Fe$_3$O$_4$(100) surfaces are expected to be similar because the computed local density of unoccupied states on Fe~\cite{PhysRevB.81.193402,SUN20111067} shows similar spin polarization at 0-1~eV above $E\rm{_F}$ on these surfaces. Second, as to the RI on surface oxygen, we need to consider the fact that oxygen atoms with and without subsurface Fe(A) neighbor exist on the surface. Since hydrogen is considered to adsorb on the latter oxygen, which is denoted as O1 in Ref.~\cite{PhysRevB.81.193402,SUN20111067}, we need to consider the change in the RI rate by H termination of O1. Following Refs.~\cite{PhysRevB.81.193402,SUN20111067}, O1 on the clean Fe$_3$O$_4$(100) has a higher DOS at around $E\rm{_F}$ due to the presence of the dangling-bond surface states while its H termination reduces the DOS greatly. However, the difference in the spin polarization at 0-1~eV above $E\rm{_F}$ between clean and H-terminated surfaces is not so clear from the previous calculations~\cite{PhysRevB.81.193402,SUN20111067}. The change in the spin dependent RI rate on oxygen by H termination of Fe$_3$O$_4$(100) is therefore is difficult to discuss.

From the above experimental observation, we expect to obtain polarized He* atoms when nonpolarized He* atoms are scattered on clean or H-terminated Fe$_3$O$_4$(100). When the incident He* beam is non-polarized, the polarization $P$ of the scattered He* is given by
\begin{equation}
\begin{split}
P=\frac{\sum_{j}(a\mathrm{_P}_{j}- a\mathrm{_{A}}_{j})b_j}{\sum_{ij}a_{ij}b_j}.
\end{split}
\end{equation}
Considering that the off-diagonal elements of the transition probability matrix are sufficiently small compared to the diagonal elements $a_{\mathrm{PP}}, a_{\mathrm{00}}$, and $a_{\mathrm{AA}}$,
we can estimate the polarization of scattered He* as
\begin{equation}
P \simeq \frac{a\mathrm{_{PP}}b_{\mathrm{P}}-a\mathrm{_{AA}} b_{\mathrm{A}}}{\sum_{j}a_{jj}b_j}.
\end{equation}
Assuming $a_{00} b_{\mathrm{0}}$ is the mean of $a_{\mathrm{PP}} b_{\mathrm{P}}$ and $a_{\mathrm{AA}} b_{\mathrm{A}}$, we obtain
\begin{equation}
P\simeq \frac{2}{3}A_{\mathrm{He*}}.
\end{equation}
The polarizations expected for scattered He* in the case of nonpolarized incident He* are therefore 0.10 for clean Fe$_3$O$_4$(100) and 0.06 for H-terminated Fe$_3$O$_4$(100). We note the following regarding the kind of surfaces preferred for the filtering. A higher filtering efficiency is expected on surfaces which have highly spin-polarized states at around $E\rm{_F}$ while their DOS is low, because both the survival probability and the spin orientation dependence of He* are expected to be high on such surfaces. Fe$_3$O$_4$ is suitable for this purpose not only because it has such properties but also because its surfaces are so inert that they would be beneficial for practical usage. Other highly spin-polarized oxides such as CrO$_2$~\cite{KSchwarz_1986} and La$_{0.7}$Sr$_{0.3}$MnO$_{3}$~\cite{PhysRevB.53.1146} may also be good for the same reason. We note that, since the He* resonance ionization happens not only on metallic cations but also on oxygen anions, the spin polarization of scattered He* gas is affected also by the local DOS and spin polarization at surface oxygen. Clean ferromagnetic surfaces such as Fe~\cite{PhysRevLett.69.2867} and Ni~\cite{PhysRevB.30.3113} are also highly spin polarized at around $E\rm{_F}$, but, since the survival probability on such surfaces would be much lower than on their oxide surfaces, they are not so appropriate for the present purpose.

\section{\label{sec:conclusions}Conclusions}
We constructed an apparatus comprising a spin-polarized He* source and a hexapole-type spin analyzer for scattered He* to evaluate the change in the spin state of He* during scattering from a surface. The experiments on clean and H-adsorbed Fe$_3$O$_4$(100) surfaces and a Cu(100) surface indicated that although the magnetic sublevel of He* is mostly preserved during the scattering, the transition between different sublevels happens with a probability of approximately 0.1 or less. We also observed the dependence of the He* survival probability on the spin orientation of He* and the Fe$_3$O$_4$ surface; this probability was associated with the spin-dependent RI rate. These findings imply that the spins of He* atoms can be filtered by scattering on ferromagnetic surfaces, depending on the local spin properties of the surface, and in particular, on the density of available levels near the Fermi energy.

The spin filtering described implies that He* should become polarized when scattered on ferromagnetic surfaces. This could be exploited to generate polarized gases with spins on solid surfaces, such as ferromagnetic and spintronic materials. Additionally, more detailed studies to elucidate the transition mechanisms of the spin states may lead to a new method for surface magnetism characterization.

\begin{acknowledgments}
This work was supported by the Sasakawa Scientific Research Grant from The Japan Science Society, National Institute for Materials Science (NIMS) Joint Research Hub Program, JSPS KAKENHI Grant No. 20H02623, and No. 20K21148, and the Doctoral Program for World-leading Innovative \& Smart Education (WISE program) of Tokyo University of Agriculture and Technology (TUAT) granted by the Ministry of Education, Culture, Sports, Science and Technology (MEXT), Japan. 
\end{acknowledgments}

\providecommand{\noopsort}[1]{}\providecommand{\singleletter}[1]{#1}%

\end{document}